%% file: mott.tex
\documentclass[aps,prl,twocolumn]{revtex4}

\usepackage{amsmath}
\usepackage{graphicx}

\def\GroupeEquations#1{\begin{subequations}  #1  \end{subequations}}
\def\moy#1{\left\langle #1 \right\rangle}
\def\Im{\hbox{Im}}

\def\SigmaSkel{{ \Sigma \kern -5.5pt \raise 1pt \hbox{/}}}

\def\sigmalatt{\Sigma^{\text{latt}}}

\begin{document}

\title{Cluster Dynamical Mean Field analysis of the Mott transition}
\author{O. Parcollet, G. Biroli  }
\affiliation{ Service de Physique Th{\'e}orique, CEA Saclay, 91191 Gif-Sur-Yvette, FRANCE  }
\author{G. Kotliar}
\affiliation{Center for Materials Theory, Department of Physics and Astronomy, Rutgers
University, Piscataway, NJ 08854, USA}
\date{ \today}

\begin{abstract}
We investigate the Mott transition using a cluster extension of
dynamical mean field theory (DMFT).
In the absence of frustration we  find no evidence for a finite
temperature Mott transition. Instead, in a frustrated model, we
observe signatures of a finite temperature Mott critical point in agreement with
experimental studies of $\kappa-$organics and with single site DMFT. 
As the Mott transition is approached, a clear momentum dependence of 
the electron lifetime develops on the Fermi surface with the 
formation of cold regions along the diagonal
direction of the Brillouin zone. Furthermore the variation of the 
effective mass is no longer equal to the inverse of the quasi particle
residue, as in DMFT, and is reduced approaching the Mott transition.

\end{abstract}

\maketitle 

\begin{figure}[hbt]
\includegraphics[width=3cm]{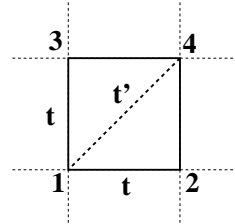}
\caption{The real space cluster and labeling conventions.}
\label{cluster.eps}
\end{figure}

The Mott transition  is one of the central issues
in the physics of strongly correlated electron systems.
Great theoretical  progress in this area  has been achieved using DMFT
\cite{DMFT}.
In spite of these successes, the single
site {DMFT} approach  has several limitations due to the
$k$-independence of the self-energy:  variations of the
quasiparticle residue, the quasiparticle lifetime and the
effective mass along the Fermi surface are absent. Moreover, some
orders involving several sites like $d$-wave superconductivity
can not be described. 
Furthermore, single site DMFT does not capture 
the effects of the magnetic exchange interaction on the single
particle properties in the paramagnetic phase. This  is
potentially a singular perturbation, which can bring
substantially new effects.  
Hence  the question
of whether the  Mott transition survives beyond the single site mean field approximation
and how it is modified by short range magnetic correlations
 is among the most pressing 
questions in the  field.\\
These questions can now be addressed 
using 
cluster extensions of DMFT  \cite{DCA-all,CDMFT-all}.
In this paper, we report a study of
the Cellular DMFT (CDMFT) approximation \cite{CDMFT-all} 
on a $2\times 2$ square cluster of the
one-band Hubbard model on a square lattice with and without
frustration, solving the CDMFT equations with the Quantum Monte
Carlo (QMC) method.  In the unfrustrated model, the 
finite temperature Mott transition is precluded by antiferromagnetic (AF) order. However, if the
frustration is big enough to destroy this order,
we show the existence of a Mott transition 
which is in agreement with the DMFT scenario and displays, at the same
time, new physical effects due to 
the $k$-dependence of the self-energy in the cluster method.
In particular, we observe the appearance of different behavior of
the one electron spectral function in different regions  of the
Fermi surface. This behavior, out of reach of single site DMFT, 
is very important since observed in the photoemission
experiments \cite{ding-hot-cold,loesser-hot-cold}  of
cuprate superconductors.\\
The one-band Hubbard model on a two-dimensional square lattice with
hopping $t_{i,j}$ and Hubbard repulsion $U$ reads :
\begin{equation}
H=-\sum_{i,j,\sigma }t_{i,j}c_{i,\sigma }^{\dagger }c_{j,\sigma
}+\sum_{i}U \biggl ( n_{i\uparrow } - \frac{1}{2}  \biggr )
\biggl ( n_{i\downarrow }  - \frac{1}{2}  \biggr )
\label{hamiltonianBasic}
\end{equation}%
where $\sigma =\uparrow ,\downarrow $ is the spin index, and  $c_{i,\sigma
}^{\dagger }$ and $c_{j,\sigma }$ denote  the electron operators.
The bare one electron dispersion  is 
given by $\ E_{k}=-2t(\cos (k_{x})+\cos (k_{y}))-2t^{\prime }\cos (k_{x}+k_{y})$.
We analyze both the unfrustrated model $t'=0$ and the strongly frustrated one
$t'\approx t$, where the antiferromagnetic (AF) order is expected to
be destroyed by frustration \cite{Marston}.
We take $D\equiv 4t=1$.
This type of anisotropic band structure is
particularly relevant for the study of organic materials like $\kappa$-(BEDT-TTF)$_{2}X$ (see e.g. \cite{Kuroki}).\\
Within CDMFT we divide the lattice in a superlattice of two by two squares
and we basically solve the \textit{DMFT}\ equations applied to the
superlattice \cite{CDMFT-all} (See Fig. \ref{cluster.eps}).
Starting from an effective action containing a Weiss dynamical field $G_{0}$
one computes by QMC (Hirsch-Fye algorithm) a local cluster propagator
and a cluster self-energy (which are four by four matrices). The Weiss
field  is then recomputed using the Dyson equation
\GroupeEquations{\label{CDMFT}
  \begin{align}\nonumber
 G_{0}^{-1} (i\omega_n ) &= \left(\sum_{K} \frac{1}{i\omega_n + \mu
 - \hat{t} (K) - \Sigma_{}(i\omega_n ) } \right)^{-1}  \!\! +
\Sigma_{} (i\omega_n )
    \end{align}
}
where $K$ is in the reduced Brillouin zone of the superlattice, $\hat t$ is the
hopping matrix for the superlattice and $\mu $ is the chemical potential. This procedure is iterated until
convergence is reached.
In our QMC runs we use $30000-300000$ QMC sweeps 
and a number of time slices $L\approx \beta U$. Finally, let us recall
that in CDMFT, the lattice self-energy is different from the cluster
self-energy $\Sigma$ and reads 
\cite{CDMFT-all} here ($k$ is in the Brillouin Zone of the square
lattice) :
\begin{multline*}
\sigmalatt (k)=
\frac{1}{4} \sum_{i=1}^4 \Sigma_{ii} +
\frac{1}{2}
\biggl [
\bigl(\Sigma_{12}+\Sigma_{34}\bigr)\cos (
k_{x})
+\\
\bigl(\Sigma_{24}+\Sigma_{13}\bigr)\cos (
k_{y})
+\Sigma_{14}\cos (k_{x}+k_{y}) +\Sigma_{23}\cos (k_{x}-k_{y})
\biggr ]
\end{multline*}
Let us first focus on the unfrustrated  case ($t'=0$) at half-filling
(See also \cite{JarrellAF,Tremblay,Haujle,Stanescu}).
We computed the N{\'e}el temperature, corresponding to the continuous onset of an
antiferromagnetic solution, as a function of $U$. Our result is similar, even
quantitatively, to the one obtained in a DMFT study of the bipartite
half-filled Bethe lattice \cite{DMFT}. In particular the AF transition
takes place at a quite high temperature, precluding a finite
temperature DMFT-like Mott transition ({\it i.e.} a first order
transition ending at a finite temperature critical point).

Let us emphasize a major difference between single-site and cluster
methods. In DMFT,  the paramagnetic 
equations ({\it i.e.} the equations with enforced spin symmetry)
 are the same for the bipartite half-filled Bethe
lattice and for a fully frustrated infinite dimensional model
\cite{DMFT}. Therefore the study of the Mott transition can be
performed analyzing the paramagnetic solution in the AF part of
the phase diagram. On the contrary, in cluster methods, the
paramagnetic equations are not the equations of such a simple
fully frustrated model. Because of the quantum
short-range fluctuations existing inside the cluster, studying
the paramagnetic solution  for parameters where an ordered
solution exists is unphysical.  Indeed if one forces the symmetry
and follow the paramagnetic CDMFT solution inside the AF region a
pseudo-gap opens because of the Slater mechanism.
Some previous work have focused on increasing the cluster size
\cite{JarrellAF,Tremblay}. The N{\'e}el temperature then 
decreases to 0, but the $T=0$ AF order can still induce long range fluctuations at finite
temperature and open a pseudo-gap.

\begin{figure}[hbt]
\includegraphics[width=7cm]{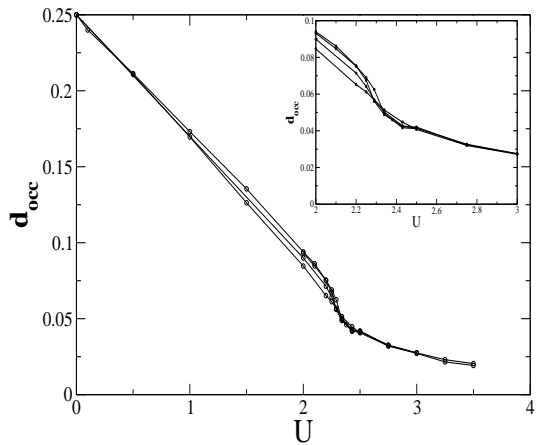}
\caption{Double occupancy as a function of $U$. The curves correspond
(from bottom to top) to $T/D =1/20,1/30,1/40,1/44$.}
\label{d.eps}
\end{figure}

\begin{figure}[thb]
\[
\includegraphics[width=8cm]{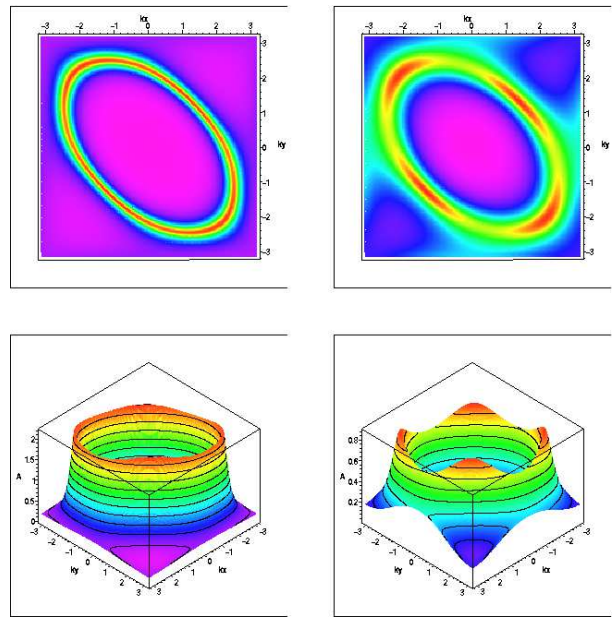}
\]
\caption{
Distribution of low energy spectral weight in $k-$space $A (k,\omega
=0)$ for $T/D= 1/44$, $U/D = 2.0$ (left
panel) and $U/D = 2.25$ (right panel). The top panel are color plots to
see the Fermi surface (red is high, blue is small) and bottom panel
are 3d plots to see the variation of $A$.
For intermediate $U$, cold and hot regions are visible around $(
\frac{\pi }{2}, \frac{\pi}{2})$ and $ (\pi, 0)$ respectively.
\label{fermi.eps}
}
\end{figure}

\begin{figure}[bth]
\vskip 15pt
\includegraphics[width=7cm]{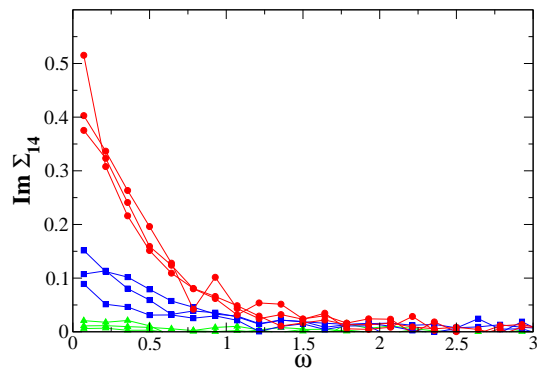}
\caption{Imaginary part of $\Sigma_{14}$ for $T/D =1/44$ and $U=2,2.1,2.2$ (green
curves), $U = 2.25,2.29,2.31$ (blue curves), $U= 2.34,2.43,2.5$ (red curves)
and $T/D =1/44$.
\label{Im2.eps}}
\end{figure}

On the contrary, in this paper we analyze a  {\it strongly frustrated
} Hubbard model with a $2\times 2$ cluster. 
In the following, we will take $t'/t=1$.
This frustration is big enough to destroy the AF order, at
least for intermediate temperatures $T/D > 0.02$ where we have
performed QMC computations.
In fact we observe unambiguous signatures of a
finite temperature Mott transition similar to
the single site DMFT case.
\begin{figure}[thb]
\includegraphics[width=8cm]{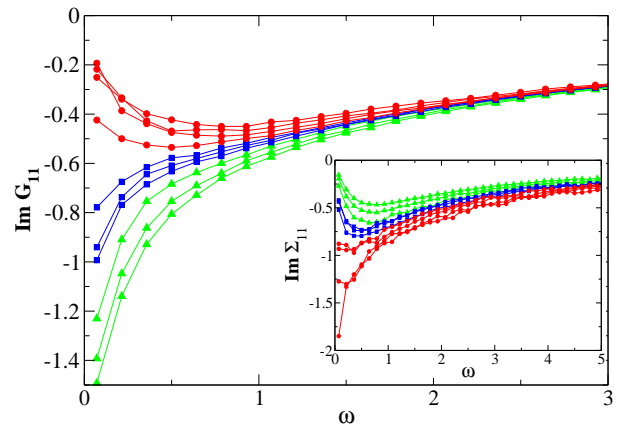}
\caption{Imaginary part of the on-site propagator and of the on-site
self-energy (inset) in Matsubara frequencies for $U=2,2.1,2.2$ (green
curves with triangles), $U = 2.25,2.29,2.31$ (blue curves with
squares), $U= 2.34,2.43,2.5$ (red curves with circles)
and $T/D =1/44$.}
\label{ImG_Sig.eps}
\end{figure}
In Fig. \ref{ImG_Sig.eps} we present the imaginary parts of the on-site cluster
self-energy and propagator $\Sigma_{11}''$ and $G_{11}''$, for different values of $U$ and as a function of the
Matsubara frequencies. For $U<2.2$ and small $\omega_{n}$, we observe
a metallic behavior: $\Sigma_{11}'' (i \omega_{n}) \sim  \left(1 - \frac{1}{Z} \right)  \omega_{n}$.
When $\omega_{n}\rightarrow 0$, the imaginary part of the on-site self-energy increases linearly, while
$G_{11}''$ decreases to a constant. 
This is the behavior characteristic of a Fermi liquid with a decreasing quasi-particle
residue $Z$  and a finite density of states at zero
frequency.
For $U>2.34$ the behavior is clearly different:
when $\omega_{n}$ decreases to 0, $\Sigma _{11}''$ decreases while $G_{11}''$
increases. This is characteristic of an insulating phase with a
(pseudo-)\allowbreak gap in the
density of states.
These results shows clearly the crossover from a
metallic to an insulating regime, similarly to DMFT \cite{DMFT}.
Furthermore, we have found indications of the existence of a finite
temperature Mott critical point, located around $U_{c}/D\simeq 2.3-2.4$ and $T_{c}/D\simeq 1/44$.
In Fig. \ref{d.eps}, we present the double occupation $d_{occ} \equiv \frac{1}{4}\sum_{i=0}^4\moy{n_{i\uparrow}n_{i\downarrow}}$
as a function of $U$ for various temperatures.  It presents a
singular behavior similar to the one found in
\cite{RozenbergKotliar} for DMFT. Furthermore the solution of the
CDMFT equations displays a numerical ``critical slowing down''
around $U_{c},T_{c}$. This is natural around a second order
transition where new solutions appears continuously.\\
Let us now turn to the new aspects of the Mott transition that one can
get with cluster methods and that are excluded from the beginning
within a DMFT analysis. As discussed previously, they are encoded into the $k-$dependence of the lattice self-energy.
First, let us focus on the Fermi surface that can be located analyzing the maxima,
as a function of $k$, of the spectral weight at zero frequency $A (k,\omega  = 0) \approx \frac{1}{\pi}\lim_{\omega_{n}  \rightarrow
0}\Im G (k, i\omega_{n})$.
Since the QMC method produces {\it finite temperature} data in imaginary
time, the self-energy at zero frequency is estimated by a linear
extrapolation  toward zero from the first two Matsubara frequencies.
In Fig.~\ref{fermi.eps}, we present $A (k,\omega  = 0)$ for two
metallic cases $U = 2.0,2.25$. The Fermi surface is only weakly
renormalized compared to the $U=0$ value.
Furthermore, 
the right part of Fig.~\ref{fermi.eps} shows that as the Mott transition
is approached the regions along the diagonal
are characterized by a larger spectral intensity, thus a smaller
electronic  lifetime, than along the $(0,\pi)$
directions. Note that the existence of hot and cold spots along the
Fermi surface have been the subject of intensive discussions in the cuprate
literature\cite{hot-cold-theory}.\\
The mechanism for the momentum differentiation
in our CDMFT study is as follows.  When $U$ increases, three sets of curves can
be distinguished: for $U=2,2.1,2.2$ (Green curves on Fig.
\ref{ImG_Sig.eps},\ref{Im2.eps}), the on-site lattice self-energy
exhibits metallic behavior, while the off-diagonal elements are
very weak, leading to a DMFT-like result and to the $A (k,\omega  =
0)$ plotted on the left in Fig. \ref{fermi.eps}. For $U=2.25,2.29,2.31$
(Blue curves), $\Sigma_{11}$  is still metallic-like  but the
$\Sigma_{14}'', \Sigma_{23}''$ are not negligible (see Fig.
\ref{Im2.eps}). We will refer to these points as the
``intermediate region''. For the largest $U$ ($2.34,2.43,2.5$,
red curves), the system is a finite temperature Mott insulator.
In this regime the real part of $\Sigma_{14} $ and $\Sigma_{23}$ 
become large and have  a sign 
such that it tends to restore the square symmetry of the lattice.
In the intermediate region, $\Sigma_{14}'' $ and $\Sigma_{23}''$
become large and of  comparable magnitude.  This is a remarkable
and unexpected effect,  since  the bare  hopping amplitude between
the 1,4 and the 2,3 sites are very different (0. and .25 ) in our
model. Hence, the anisotropy of the electron lifetime stems from
$\Sigma_{14}'' \cos (k_{x}) \cos(k_{y})$.
This quantity is maximum and negative around
$(0,\pi)$ and $(\pi,0)$ (hot spots) and is a minimum, i.e. zero, at
the diagonal $({\pi /2}, {\pi / 2})$ (cold spots) and leads to
the curves plotted on the right in Fig. \ref{fermi.eps}. 
Note however that this is an {\it intermediate temperature} effect.
Indeed a scenario where, at zero
temperature and $\omega \rightarrow 0$, $\Sigma_{11}'' \rightarrow 0$ and $\Sigma_{14}'' \sim const$ (this would lead to a
non Fermi liquid behavior more pronounced around $(0,\pi)$ and
$(\pi,0)$) is 
forbidden by causality (and thus by CDMFT): $\Sigma''$ has to be
negative as a matrix for all $\omega$, but if $\Sigma_{11}'' ( \omega
=0) = 0 $  and $\Sigma_{14}'' (\omega =0) <0 $ then
the submatrix $\Sigma_{ij}$ with $i,j \in \{1,4 \}$ would not be
negative.

\begin{figure}[thb]
\includegraphics[width=7cm]{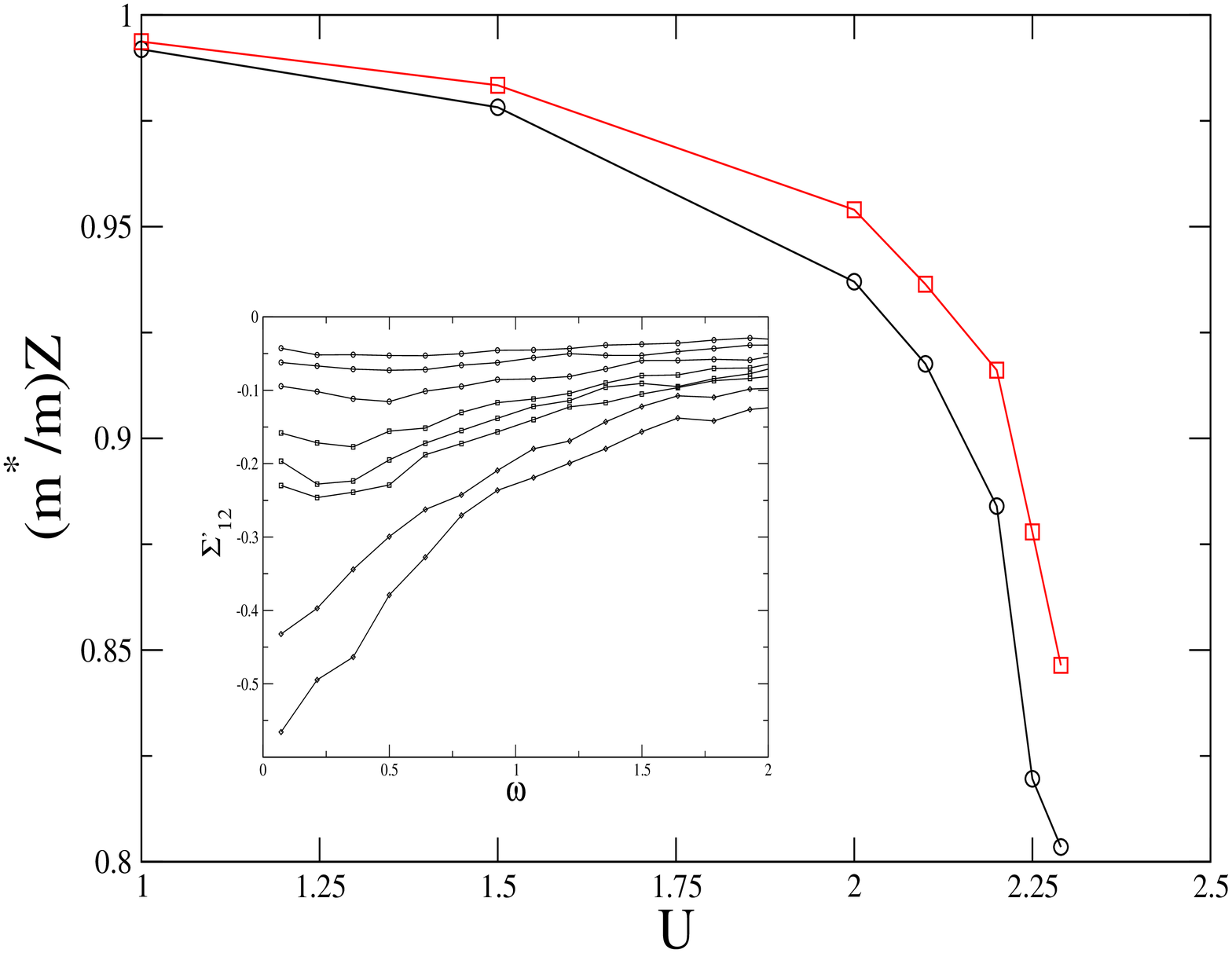}
\caption{$Z\frac{m^{*}}{m}$ as a function of U at $\beta D =44$. Top
(bottom) curve corresponds to the hot (cold) point.
Inset :$\Sigma'_{12}$ vs. $\omega_{n}$ for
$U=2,2.1,2.2,2.25,2.29,2.31$ (from top to bottom) and $T/D=1/44$.}
\label{m.eps}
\end{figure}

More refined quantities are the variation of the quasi-particle
residue and the effective mass on the Fermi surface. We
find that quasi particle residue 
decreases with $U$, roughly linearly and becomes
small near $U\approx U_{c}$ as in DMFT. However, since it involves
the derivatives of the  analytically continued self-energy, it can
not be evaluated within our QMC results with enough precision to
differentiate the behavior of $Z$ in the hot and cold region. On
the other hand, an interesting physical quantity, that can be reliably 
extract from the real part of the self energy at zero frequency, is:
\[
Z\frac{m^{*}}{m}  =\frac{1}{1+\frac{d \Sigma' (k,0)}{dk_{\perp}
}/\frac{dE_{k}}{dk_{\perp}}}
\]
where $m^{*}$ and $m$ are respectively the renormalized and the
bare effective mass, $k$ is along the Fermi surface 
and $\frac{d}{dk_{\perp}}$ means the derivative perpendicular to the Fermi
surface. $\frac{m^{*}}{m}Z$ is plotted in Fig. \ref{m.eps} 
for $\beta D =44$ as a function of $U$ for the hot
and cold points. Note that $\frac{m^{*}}{m}Z$ is always equal 
to one in single site DMFT because of the $k$-independence of the self
energy. Instead within CDMFT we find that it decreases as we approach the
Mott transition mainly because of the increasing in modulus of 
the real part of $\Sigma_{12}$ at zero frequency, see inset of Fig. \ref{m.eps}.
This leads to a contribution to the $k$-dependent
renormalization of the effective mass which tends to counterbalance
the increase of $1/Z$. Whether the
effective mass at low temperatures remains finite as the Mott point
is approached as in the large $N$ studies of Ref \cite{grilli}, or
whether it diverges in an albeit weaker fashion than in DMFT,
remains an open question which will require a lower temperature
study.\\
Finally, analyzing the density of states using the Maximum Entropy
\cite{MaxEnt} analytic continuation procedure we have found that
the quasi particle peak found in DMFT in the metallic region is
split near the Mott transition into two peaks, giving rise to a
pseudo-gap \cite{workinprogress}.
 
To summarize, we found that the Mott transition, prevented by AF order
in the unfrustrated model, is present in the frustrated case.
The single site DMFT scenario characterized by a finite temperature
Mott endpoint is compatible with our data. At low temperature
a substantial $k$-dependence of the self-energy shows up and 
regions with significant variation of the electron lifetime 
as a function of $k$ appear. 
Our analysis has been performed on a model
relevant to the quasi-two
dimensional materials of the $\kappa$ family \cite{organics0,organics1}. Indeed in Ref \cite{organics1} the Mott
transition has been recently driven by pressure. 
It would be very interesting to carry out  angle resolved photoemission
experiments  in these materials to verify our theoretical
predictions, regarding the strong momentum dependence of the
lifetimes. We conjecture, that this differentiation in momentum
space is more general than the model we studied, and is direct
consequence of the proximity to the Mott transition. Indeed it would be
interesting to perform a similar study in a doped and isotropically frustrated
Hubbard model relevant for high-$T_{c}$ superconductors\cite{workinprogress}.

\begin{acknowledgments}
The computations were carried out at the parallel cluster of the  Physics Department
at Rutgers. We thank C.~Uebing for technical help with the
computers, A. Perali for his  participation  in  the early stages of
this work.

\end{acknowledgments}

\input{mott.bbl}

\end{document}

%% file: mott.bbl
\newcommand{\PRB}{Phys. Rev. B}\newcommand{\PRL}{Phys. Rev. Lett}\newcommand{\NPB}{Nucl. Phys.}\newcommand{\RMP}{Rev. Mod. Phys.}\newcommand{\ADV}{Adv. Phys.}